\documentstyle{mn}
\title[Radio galaxy unification]{Unification in the low radio luminosity regime;
evidence from optical line emission} 
\author[M.J.M.March\~a,
I.W.A.~Browne, N. Jethava, S.~Ant\'on]{M.J.M. March\~a$^{1}$
\thanks{e-mail:mmarcha@oal.ul.pt}, I.W.A. Browne$^{2}$,
N. Jethava$^{2,3}$, S.~Ant\'on$^{1}$\\ $^1$~CAAUL, Observat\'orio
Astron\'omico de Lisboa, Tapada da Ajuda, 1349-018 Lisboa, Portugal\\
$^2$~University of Manchester, Jodrell~Bank~Observatory, Macclesfield,
SK11 9DL, U.K.\\ $^3$~Chalmers University of Technology, Onsala Space
Observatory, Onsala, Sweden\\}

\begin{document}

\input{psfig.tex}

\date{may 6, 2005}

\pagerange{\pageref{firstpage}--\pageref{lastpage}} \pubyear{2004}

\maketitle

\label{firstpage}

\begin{abstract}

We address the question of whether or not the properties of all
low-luminosity flat spectrum radio sources, not just the obvious BL
Lac objects, are consistent with them being the relativistically
beamed counterparts of the low radio luminosity radio galaxies (the
Fanaroff-Riley type 1 - FR~I). We have accumulated data on a
well-defined sample of low redshift, core-dominated, radio sources all
of which have one-sided core-jet structures seen with VLBI, just like
most BL Lac objects. We first compare the emission-line luminosities
of the sample of core-dominated radio sources with a matched sample of
FR~I radio galaxies. The emission lines in the core-dominated objects
are on average significantly more luminous than those in the
comparison sample, inconsistent with the simplest unified models in
which there is no orientation dependence of the line emission. We then
compare the properties of our core-dominated sample with those of a
sample of radio-emitting UGC galaxies selected without bias to core
strength. The core-dominated objects fit well on the UGC correlation
between line emission and radio core strength found by Verdoes Kleijn
et al. (2002). The results are not consistent with all the objects
participating in a simple unified model in which the observed line
emission is orientation independent, though they could fit a single,
unified model provided that some FR~I radio galaxies have emission
line regions which become more visible when viewed along the jet axis.
However, they are equally consistent with a scenario in which, for the
majority of objects, beaming has minimal effect on the observed core
luminosities of a large fraction of the FR~I population and that
intrinsically stronger cores simply give rise to stronger emission
lines. We conclude that FR~I unification is much more complex than
usually portrayed, and models combining beaming with an intrinsic
relationship between core and emission line strengths need to be
explored.

\end{abstract}


\section{Introduction}

The utility of unified schemes of active galaxies is their simplicity
and their potential for making testable predictions. In the case of
powerful FR~I radio sources some of the initial simplicity has been lost
because modifications have had to be introduced in order to match the
predictions to the observations (see Urry \& Padovani, 1995 for a
review). In the case of low power radio sources, however, the simple
idea that FR~I radio galaxies looked at down their jet axes become BL
Lac objects has remained the consensus view (Padovani \& Urry,
2001). This is perhaps somewhat surprising since only about a third of
low-luminosity core-dominated radio sources (supposedly the beamed
counterparts of FR~Is) are found to be conventional BL Lac objects; most
of the remaining two-thirds of such objects have optical
classifications such as Seyfert-like objects or passive elliptical
galaxies (March\~a, et al., 1996). What is the relationship between
these other types of core-dominated objects and FR~Is?

We have been exploring the idea that the synchrotron cores of all low
luminosity core-dominated radio sources are intrinsically similar,
irrespective of their emission line properties, and that all might be
appropriately labeled blazars. Such a scheme is consistent with their
continuum SEDs (Caccianiga \& March\~a, 2004; Ant\'on et al., 2004;
Ant\'on \& Browne, 2005).

In this paper we investigate the emission-line properties of low
luminosity radio galaxies, both core-dominated and lobe-dominated. We
first compare the emission-line properties of a group of
core-dominated objects with those of a matched sample of FR~I radio
galaxies. According to the simplest unified scheme, in which there is
no orientation dependence of the line emission, the emission line
properties of the two samples should be statistically
indistinguishable. A rigorous exploration is possible because we now
have available a sample of core-dominated radio sources - the 200~mJy
sample (March\~a et al., 1996) - for which we have emission line
luminosities, extended radio luminosities and high-resolution VLBI
observations. Knowing the extended radio luminosities is potentially
important because the emission-line properties of AGN correlate with
radio power (Rawlings \& Saunders, 1991; Verdoes Kleijn et al., 2002)
and it is therefore necessary to choose a comparison sample from
objects having the same range of intrinsic radio powers. From a
unified model point of view, the weak extended radio emission of the
core-dominated objects (i.e. excluding the core and one-sided jet
emission) should be a good measure of the intrinsic radio power since
the symmetry of the emission, and its low surface brightness,
eliminates the possibility that its observed strength is influenced by
relativistic beaming. The high-resolution VLBI observations are also
very important since we cannot  simply rely on spectral selection to pick
core-dominated sources since not all radio sources with flat spectra
have radio structures dominated by cores. Some are found to be Compact
Symmetric Objects (CSOs) in which the flat spectrum does not arise
from compact self-absorbed core emission. VLBI observations enable us
to select with confidence the genuine one-sided core-jet objects and
discard the rest.

A second line of investigation can be adopted. If the non-BL Lac
core-dominated objects are not relativistically beamed, then we are
measuring the intrinsic core strengths and thus objects
with strong cores might be expected to produce more line
emission. Thus, looking at correlation between line emission and radio
core strength might be useful. Verdoes Kleijn et al. (2002) have done
this for a sample of UGC galaxies selected without reference to radio
core strength and find quite a strong correlation. From the tightness
of the correlation they conclude that the bulk Lorentz factors lie in
the range 2-5 for continuous jets, or $\leq$2 for jets consisting of
discrete blobs. Most of the objects in their sample are not
core-dominated. It would, therefore, be interesting to see if other objects
with more dominant radio cores (in which the effects of beaming would
be much more apparent if there existed bulk relativistic motions) lie
on a continuation of their correlation. We can use the data on the
200~mJy sample and combine them with the UGC results for this purpose.

The structure of the paper is as follows. In Section 2 we describe the
selection of the core-dominated sample and of a comparison sample
matched in flux density and redshift to the first. In Section
3, we briefly describe some radio observations of members of the
core-dominated sample that were made in order to determine their
extended radio flux densities. In Section 4 we compare the emission
line distributions of the two samples and also investigate where
the 200~mJy objects lie on the Verdoes Kleijn et al. (2002)
correlation. Finally, in Section 5, we discuss the results, both in
terms of unified models and in terms of ``dis-unified'' models.
Throughout we adopt a value for the Hubble constant of 75 km s$^{-1}$
Mpc$^{-1}$.

\section{The selection of the samples}
\label{samples}

We start with the 200~mJy sample (March\~a et al., 1996) updated by
Ant\'on et al. (2004). This sample consists of flat-spectrum sources
(S$\propto \nu^{-\alpha}$ with $\alpha \leq$0.5) stronger than 200~mJy
at 5~GHz, as recorded in GB6 (Gregory et al., 1996), with red optical
magnitude brighter than 17. The magnitude selection means that out to
a redshift of $\sim$0.1 we have an essentially volume-limited complete
sample. The sample is further restricted to declinations
$\geq+20^{\circ}$ and Galactic latitudes
$\geq$12$^{\circ}$. Measurements of the luminosities of the
H$_{\alpha}$ lines are available for most objects\footnote{When we
refer to H$_{\alpha}$ we implicitly include emission from the [NII]
lines with which H$_{\alpha}$ is usually blended.}. VLBI maps have
been made of virtually all the objects at 5~GHz and/or 1.6~GHz (Bondi
et al., 2001; Bondi et al., 2004; Bondi \& Polatidis, personal
communication) and we have only included in our discussion those
objects with a core and one-sided jet visible in the VLBI maps -
future maps may well reveal more core-jet objects. We have also
restricted our sample for the current paper to objects with redshifts
$\leq$0.1 since in this redshift range we are confident that the
sample is complete. The redshift restriction also facilitates finding
a good comparison sample (See below and Table 1). Spectroscopically,
the sample has been classified according to the following types (also
identified in Table 1): PEG - stands for Passive Elliptical Galaxies
and it refers to sources with weak emission lines and a strong galaxy
component; Sy1,2 - stands for sources with strong emission lines in
their spectra (1 if broad emission lines are seen, 2 if only narrow
lines are present); hyb - stands for sources with broad but
significantly weaker emission lines than in the Sy case; BL Lac -
stands for sources with weak or absent emission lines in their
spectra. Seven BL Lac objects with continuous optical spectra and no
measured redshifts from the 200~mJy sample are excluded from the
sample to be discussed. This should not bias the statistics because it
is almost certain that all these objects are at redshifts $\geq$0.1,
otherwise the host galaxy would be visible and its redshift known.

We use as a basis of our comparison sample the radio galaxies found in
Abell clusters and studied by Owen et al. (1995) and by Ledlow \& Owen
(1995). These consist of objects with 20~cm flux densities
$\geq$10~mJy and redshifts $\leq$0.09. H$_{\alpha}$ luminosities (or
limits on luminosities) are given by Ledlow \& Owen for most of the
objects. These sources are mostly ellipticals from the spectroscopic
point view, although a small number (less than 13\%) has strong
emission line spectra related to starformation (Owen et al. 1996). In
our analysis we will use a sub-sample of the Ledlow \& Owen sample
consisting of objects with H$_{\alpha}$ measurements and picked to
match the 200~mJy objects both in redshift and in flux
density\footnote{Since we have good radio spectra for the 200~mJy
sources we do the matching in extended flux density at 20~cm. The
choice of wavelength to do the matching does not affect the validity
of the comparison.} For each 200~mJy object we have looked for a
``twin'' matched to within a factor of two in 20~cm flux density to
the 200mJy object's extended flux density (see next section) and
$\sqrt{2}$ in redshift (in most cases the matches are much closer),
regardless of the strength of the line emission. Matches were possible
in all but one case. The procedure used to produce a comparison sample
eliminates any worry about the effect of the correlation between line
luminosity and the intrinsic radio power. It also ensures that the
linear scales of the spectrograph slit, when projected on the galaxy,
are comparable. The twin for each 200~mJy source is given in Table
1. In this process we are implicitly assuming that the Ledlow \& Owen
objects are representative of the population of unbeamed counterparts
of the 200~mJy objects.

\section{VLA observations and data reduction}

The best estimate we have of the intrinsic radio power of a
core-dominated object, where the core and jet emission may be Doppler
boosted, is the extended radio emission. Estimates of the extended
emission can be made using the FIRST survey (Becker et al., 1995) data
but these are available for only a fraction of the 200~mJy objects.
For this reason we have made our own VLA B-configuration observations at 20~cm
of those 200~mJy objects which do not lie in the region covered by the
FIRST survey. Each source was observed at two widely separated hour
angles for a total observing time of $\sim$10~min. The observations were
calibrated and mapped using standard AIPS tasks. The data were of good
quality and in most of the maps, extended radio structure is
detectable. The peak flux densities and total flux densities were
measured with the task IMEAN and the extended flux density taken as
the difference between the two quantities. The results are listed in
Table 1. For those sources in the FIRST survey, FITS images were
retrieved and processed in the same manner as our own maps.

\section{The emission line properties of core-dominated and lobe-dominated 
objects}

\begin{figure*}
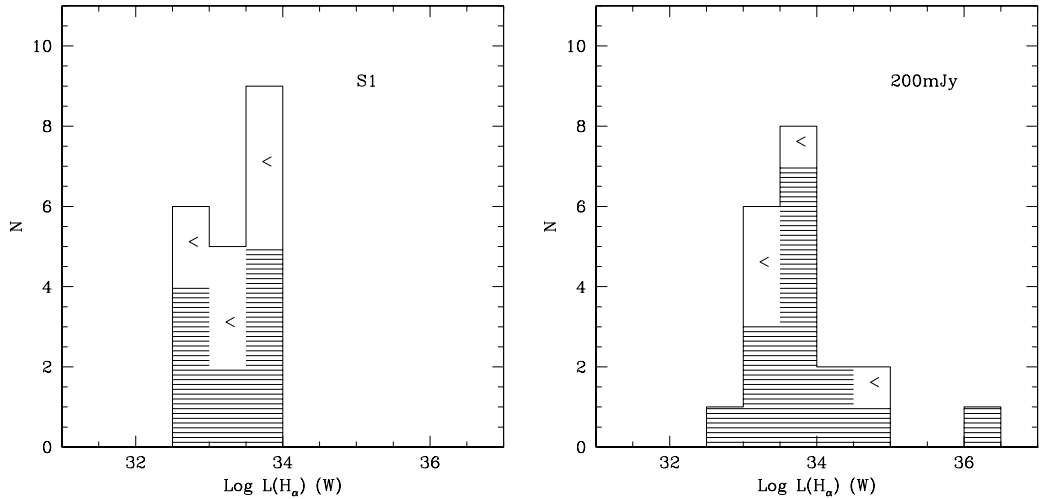

\centerline{ \psfig{file=fig.his_lha_twin,height=7cm}
\psfig{file=fig.his_lha_200,height=7cm} }
\caption{Distributions of the line luminosity for the 200~mJy sample
(left panel), and its twin sample (right panel and marked S1). 
The histograms marked with $<$, correspond to upper limits. }
\label{fig.his_avni}
\end{figure*}

\subsection{The 200~mJy sample and comparison sample}

We first ask the simple question: are the emission line properties of
the 200mJy sample sources statistically distinguishable from those of
the sources in the comparison sample? We choose to compare
H$_{\alpha}$ luminosities since these are widely available and do not
worry at this stage whether we are dealing with narrow-line or
broad-line emission. In the simplest version of the FR~I unified
scheme in which there are no hidden emission line regions one would
expect the distribution of emission line luminosities in the 200~mJy
sample and in the comparison sample to be indistinguishable (see
Figure \ref{fig.his_avni}).

An advantage of having matched pairs of objects is that it gives us a
simple way to quantify the probability that the distribution of line
luminosities for the two samples is indistinguishable.  In 16 pairs
(excluding 3 pairs where both members have limits on luminosities and
one where they were identical) it is found that in 13 cases the member
drawn from the 200~mJy sample has the more luminous H$_{\alpha}$
line. We use the binomial distribution to work out the probability
that in 13 or more cases the 200~mJy member should have the larger
line luminosity, on the assumption that the line luminosity
distributions of both groups are the same. We conclude that there is
less than 2\% probability that the line luminosity distribution in the
two samples is drawn from the same population.

The confidence in this result could be questioned due to two factors:
firstly there may be some degeneracy in the choice of twins (e.g. more
than one possible twin for each of sources of the 200~mJy sample), and
secondly, there is the statistically difficult issue of comparing
distributions where there is a significant number of upper limits. We
decide to investigate the consequences of both of these aspects by
proceeding in the following way:

\begin{itemize}

\item For each 200 mJy sample source we take all the twins that fall in the
region within a factor of two in 20cm flux density and a factor of
$\sqrt 2$ in redshift;

\item For the 16 sources with 4 twins or more we take the line
luminosities (the detections and upper limits) and find the
Kaplan-Meier estimator for the distribution function of a randomly
censored data by using survival analysis ({\em ASURV}; LaValley, Isobe
and Feigelson, 1990). For the remaining sources, we preferentially
took the twin with a detection, or in case there were only upper
limits, the highest of these as the line luminosity for the twin.

\item The line luminosity of the twins found in the way described
above is then compared to the line luminosity of the 200~mJy
sources. We find that in 17 cases the line luminosity of the 200~mJy
sources is higher than that of the twin, in 1 case it is equal, and in
2 cases it is smaller. If we then use the binomial distribution
to estimate the probability that in 17 out of the possible 19 pairs,
the line luminosity of the 200~mJy source is larger than that of the
comparison sample we find that this probability is $<$0.0001.

\item Finally, we note that 17/20 sources of the twin sample have
$L(H_{\alpha}) < 10^{33} $ W, whilst in the 200~mJy sample, only 3/20
have such low values of line luminosity.

\end{itemize}

We therefore think that the difference between the strength of the
line luminosity of 200~mJy sample and its twin should be taken
seriously. There is at least one selection effect which is likely to
mask the true difference between core- and lobe-dominated objects when
we use our two samples. The two samples are cross-contaminated; the
200mJy sample contains some not very core-dominated objects while the
Ledlow and Owen sample contains some objects which are just as
core-dominated as the 200mJy objects.  For example 0055+300 (NGC315) a
giant FR~I radio galaxy (Bridle et al., 1976) is in the 200~mJy
sample\footnote{It is expected that some radio galaxies with strong
radio cores will find their way into the 200~mJy sample simply because
of random errors on the 5~GHz and 1.4~GHz flux densities used to
define the sample}. Such a cross-contamination is likely to make it
more difficult for us to recognize true differences in emission-line
properties rather than leading to spurious differences. Hence, the
detected difference is likely to be a lower limit on the true
difference between the line luminosities of the two samples.

There is one caveat related to the fact that the Ledlow and Owen
objects are selected to lie in Abell clusters of galaxies whereas the
200~mJy sources are selected without any reference to their cluster
environments. It could be that the hosts of radio sources in clusters
have different emission-line properties to galaxies in the field. Such
an effect has been claimed to exist by Guthrie (1981). We are,
however, reassured that this may not be too important an effect since
the 200~mJy objects appear to be consistent with the correlation for
non-core-dominated objects found by Verdoes Kleijn et al. (2002)
between emission line luminosity and radio core luminosity (see below
and Figure \ref{fig.lvlbi_lha}).

\subsection{The 200~mJy sample and UGC galaxies}

The result above suggests a simple interpretation which is that
emission line strength is related to core strength.  Such a view is
supported by the observation of a tight correlation of the core
emission line strength with radio core strength in the UGC FR~I
radio-galaxies (Verdoes Kleijn et al., 2002; Xu et al., 2000; Figure
2).  Given that most of the residual dispersion in the correlation
could be due to measurement errors and/or to radio core variability,
that leaves little room for the effects of beaming on the core
strengths. Variability is relevant for narrow-line emission because
the Narrow Line Region (NLR) probably has a size of $\sim$tens to
hundreds of parsecs and thus its observed strength reflects the core
activity integrated over times-scales of tens, perhaps hundreds, of
years. Hence, one spot measurement of radio core strength may not be a
good indicator of the integrated core activity.

\begin{figure}
\centerline{
\psfig{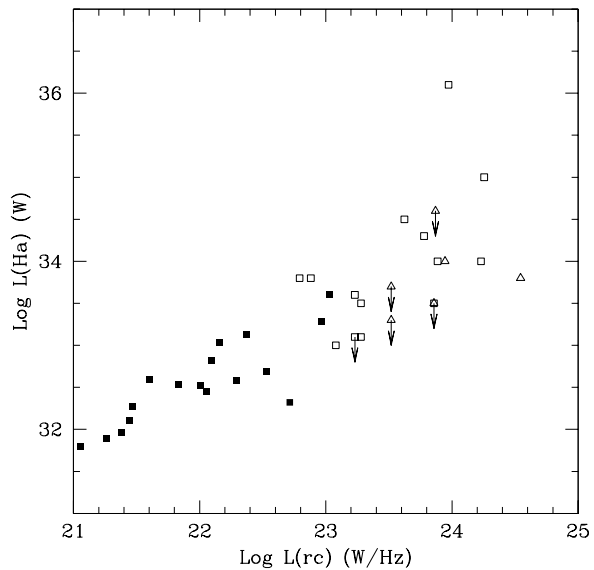} 
}
\caption{Distribution of the line luminosity vs. the radio core
luminosity for the UGC FR~Is of Verdoes Kleijn et al. (filled squares)
the 200~mJy sample (open symbols; triangles are the BL Lacs). 
}
\label{fig.lvlbi_lha}
\end{figure}

To investigate if the correlation holds for objects contained in the
200~mJy sample, we include these sources with the UGC objects of
Verdoes Kleijn et al.  in a plot of the VLBI core luminosity ($L_{rc}$)
against the line luminosity $L(H\alpha)$ (Figure \ref{fig.lvlbi_lha},
where we separate the BL Lacs in the 200~mJy from the remaining
sources by plotting them as open triangles). Because the 200~mJy
sources have more dominant cores than the UGC objects, they might be
expected to display the effects of beaming more strongly. (We note
that the comparison sample is not included in this plot since they lack
VLBI measurements). This should manifest itself in two ways: (i) as an
increase in the dispersion in the correlation, and (ii) in the more
core-dominated objects falling on average below the line established
for the UGC objects. \footnote{We note that the plotted line
luminosities are slightly different in the two samples. In the case of
the UGC objects the line emission refers to the nuclear
H$_{\alpha}+$[NII] emission measured with HST, whereas in the case of
the 200~mJy objects we are using the H$_{\alpha}+$[NII] flux
integrated over a wider spatial extent.} It is clear that the 200~mJy
objects join relatively smoothly on to the existing UGC correlation.

The correlation coefficient for the UGC+200~mJy objects (excluding the
6 BL Lacs in the 200~mJy where 4 have upper limits to the line
luminosity) is 0.72 corresponding to a probability of there being no
correlation of $\ll$1\%. The linear least squares fit gives a slope of
0.89$\pm0.1$. There is an increase in
the scatter going from the UGC to the 200~mJy objects but no obvious
sign in the decrease in slope which would be expected if the radio
core emission is beamed and the line emission isotropic. In fact, the
BL Lacs of the 200~mJy sample seem to fit within the
general relationship without significantly increasing the dispersion
of the distribution.  This is confirmed by the generalised Spearman's
correlation test of 0.76, when we use survival analysis to include the
upper limits of the line luminosity of the BL Lacs. This value means
that the probability of no correlation between the two quantities is
less than 1\%. The resulting linear regression when all the sources
are considered is $L(H_{\alpha}) \propto (0.76 \pm 0.09) L_{rc}$.

We emphasise that the correlation between radio core and line
luminosity is not induced by redshift. In fact, the partial
correlation coefficient between the two quantities, excluding the
effect of redshift is 0.62 for the UGC and 200~mJy sources (where the
six BL Lacs were not considered) which gives a probability $<< 1$\% to
the hypothesis that there is no correlation between the two
luminosities.

\section{Discussion}

The basis of our discussion is that the more core-dominated sources
have significantly stronger line emission than the lobe-dominated
sources. This shows up in the comparison of the 200~mJy sample objects
with the matched sample objects and in the strong correlation between
radio core luminosity and emission line luminosity. What, if anything,
does this tells us about FR~I unification? We will start by looking at
things from two opposite viewpoints, one unified and one
``dis-unified''. We then discuss if there is a middle way.

\subsection{The unified interpretation}

If we adopt a strict ``orientation is everything'' point of view, we
conclude that in some FR~Is, optical emission line regions must often
be hidden from the observer, and (statistically) become more visible
when looking close to the jet axis. We note that in most cases, but
not all, the line emission we see in the core-dominated objects of the
200~mJy sample is not nuclear broad-line emission (which is believed
to be present in most FR~IIs but sometimes hidden from view by a
molecular torus), but most often narrow-line emission which originates
at radii more than a few parsecs from the nucleus of the galaxy. More
specifically, there are four objects, 0125+487, 0321+340, 1646+499 and
2116+81, in the 200mJy sample in Table 1 which are classified as
having broad emission lines and all four have high line
luminosities. Although better (higher resolution) spectra would be
required to provide narrow line luminosities, visual inspection of the
available spectra shows that all the broad line sources have
substantial narrow line components which are probably sufficiently
strong by themselves for the galaxies to be called Seyfert 2s. This
means that in nearly all cases it is the narrow line emission, which
is thought to originate more than a few parsecs from the nucleus, that
would have to be hidden. This type of narrow-line emission would not
be affected by a torus but might be partially hidden by, for example,
extinction in a kind of dust and gas disks seen in many FR~I radio
galaxies (de Ruiter et al., 2002)\footnote{We note that Chiaberge et
al. (2002) argue that the combination of optical and UV HST data, and
radio core emission for FRIs is not consistent with the standard thick
torus framework.}.  Additional evidence for off-nuclear gas and dust
is presented by Quillen et al. (2003) who argued, on the basis of 3CRR
radio-galaxy Spectral Energy Distributions (SEDs), that the nuclear
regions of some FR~I radio galaxies suffer from optical
extinction. Further evidence comes from recent results by Wills et
al. (2004) on 13 low luminosity sources of the 2~Jy sample. In
particular, these authors found that the average [OIII] line
luminosity in BL Lacs is significantly larger than that of FR~Is, thus
supporting the view of extinction in the latter type of sources. Since
a factor of a few in orientation-dependent extinction in the inner kpc
would be enough to account for the excess of line emission in the
core-dominated objects with respect to those in the comparison sample,
we think this is just about a viable scenario.

\subsection{A dis-unified interpretation}

\begin{figure}
\centerline{
\psfig{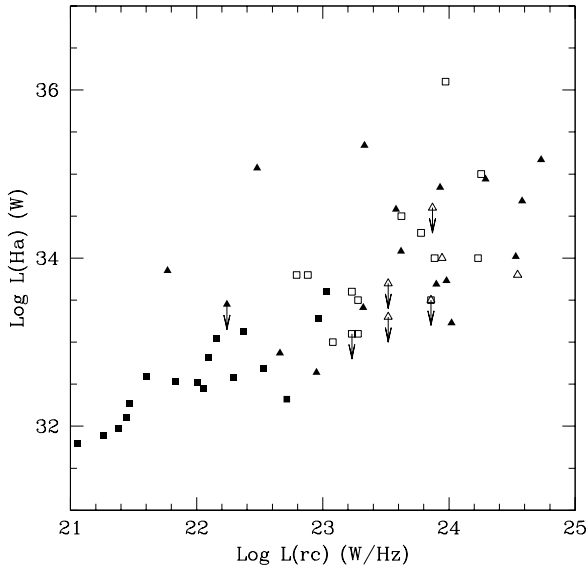} 
}
\caption{Distribution of the line luminosity vs. the radio core
luminosity for the UGC FR~Is of Verdoes Kleijn et al. (filled squares)
the 200~mJy sample (open symbols; triangles are the BL Lacs), and the
3CR FR~1 galaxies of Cao \& Rawlings (filled triangles). The
radio core flux densities were converted to 1.4~GHz using a radio
spectral index $\alpha_{r}=0$}
\label{fig.lvlbi_lha2}
\end{figure}

An alternative interpretation is to say that the observations rule out
FR~I unification or, at least, force us to exclude from any such
schemes the core-dominated synchrotron jet sources which are not
recognizably BL Lacs. This would not be a very tidy state of affairs,
but it could be the truth!  Such a view, one might argue, is the most
logical interpretation of the tight correlation of the core emission
line strength with radio core strength in the UGC FR~I radio-galaxies
(Verdoes Kleijn et al., 2002),  and the fact that the 200~mJy objects
lie on the same correlation (Figure 2). To explore this correlation
further we have added to the objects plotted in Figure 2 those FR1
radio sources from the 3CR studied in Cao \& Rawlings (2004) - see
Figure \ref{fig.lvlbi_lha2} where we have excluded the 3CR FR~I
galaxies that were in common with those already present in the UGC
sample. We emphasize that no attempt has been made to match the
200~mJy sample objects in extended luminosity. It is clear that the
core luminosities cover a similar range to those seen in the 200~mJy
sample and the line luminosities are consistent with objects from this
additional sample participating in the same correlation as the objects
from the other two samples. Simple beaming schemes, even those
incorporating orientation dependent extinction of emission line
regions would not predict this. 

\begin{figure}
\centerline{
\psfig{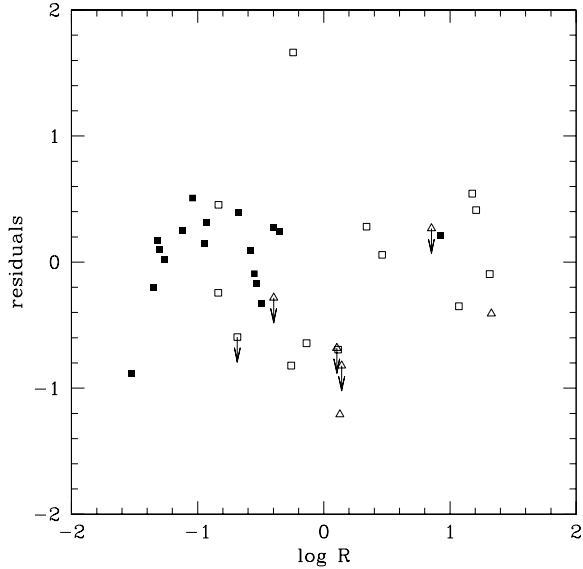} 
}
\caption{Distribution of the residuals of the line luminosity - radio
core luminosity correlation (see text for details) against the radio
core parameter $R$ for the UGC FR~Is of Verdoes Kleijn et al. (filled
points) and for the 200~mJy sample (empty symbols; triangles are BL
Lacs)}
\label{fig.r_res}
\end{figure}

\begin{figure}
\centerline{
\psfig{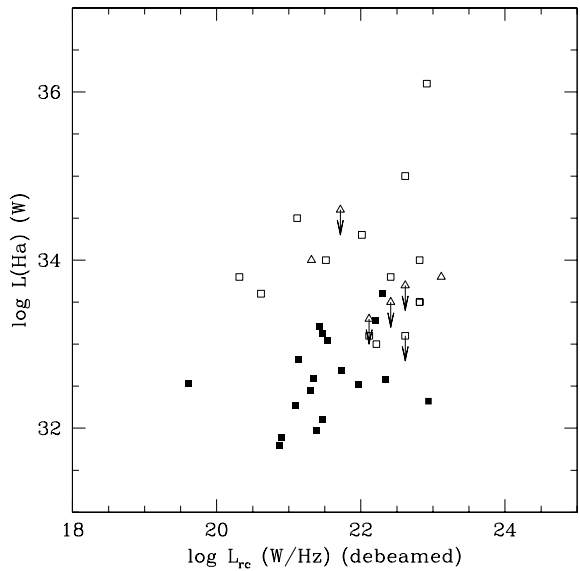} 
}
\caption{Distribution of the line luminosity vs. the radio core
luminosity for the UGC FR~Is of Verdoes Kleijn et al. (filled points)
and for the 200~mJy sample (open symbols) after the core luminosity
has been 'de-boosted' (see text for details).}
\label{fig.lrc_lha_debeamed}
\end{figure}

One might argue that because the slope of the core/emission line
correlation is less than unity, a viable hypothesis is that there is
an underlying linear correlation between H$_{\alpha}$ luminosity and
intrinsic core strength, and all the scatter 
in Figure 2 could be attributable to the effects of beaming. In
other words, beaming should leave H$_{\alpha}$ unchanged but move the
core luminosity towards the right  in Figure 2. We have
tested this idea in two ways. First we look for a correlation between
the magnitude of the deviation of the core luminosity from the
underlying linear (slope unity) correlation and the core dominance
parameter R ($L_{rc}/L_{ext}$ at 1.4~GHz). Both should be orientation
indicators in the beaming scenario. We find absolutely no correlation
(see Figure \ref{fig.r_res}), which reinforces the view that the
evidence for radio core beaming effects amongst these objects is very
weak. The second approach is to adopt the traditional unified scheme
view (e.g. Orr \& Browne, 1983) that there is a tight relation between
the intrinsic core strength and that of the extended lobe emission. In
this case, the measured core-dominance (R) compared to what would be
seen if the object were viewed in the plane of the sky (R$_{c}$) is a
direct measure of the core Doppler boosting. Based on models which use
population statistics to constrain beaming model parameters, current
estimates for R$_{c}$ are about 0.01 (Wall and Jackson, 1997; Jamrozy,
2004).  We use these estimates, and the observed R values, to de-boost
each object. The resulting plot of line luminosity against de-boosted
core luminosity is shown in Figure \ref{fig.lrc_lha_debeamed} where
symbols are as before. The original correlation is
destroyed. 

One additional thing we have tried, still within the context of there
being a linear correlation between line luminosity and intrinsic core
strength, is to estimate how much orientation dependent extinction
would be required to just counterbalance the effect of beaming and
restore Figure 5 to look like Figure 3. The result of our exercise,
however, is that it would require one to move most of the 200~mJy
objects in Figure 5 by around two orders of magnitude in line
luminosity which is much larger than current estimates of the
extinction suffered by the cores of most FR~I radio galaxies would
allow. For instance, based on the study of the UV emission of the
nuclei of 3CR galaxies, Chiaberge et al. (2002) deduce a median value
of $A_{V}=1.3$ for the FR~Is in their sample.

\subsection{Reconciling the interpretations}
 
We are faced with a dilemma. The evidence for relativistic bulk motion
in the cores of BL Lacs is extremely strong and hence the core
emission we observe in these must be highly orientation dependent
because of Doppler boosting. There is also direct evidence from
superluminal motion that there are highly relativistic jets at the
pc-scale in FR1 radio galaxies (Giovannini et al., 2001). The case for
unification between BL Lacs and some, or all, FR1s is thus very
strong. On the other hand, we have shown that BL Lacs are not the only
low-luminosity core-dominated radio sources with synchrotron jets
(March\~a et al, 1996; Bondi et al., 2001; Ant\'on et al., 2004, Bondi
et al., 2004), and is very tempting to hypothesise that these non-BL
Lacs have relativistic jets too and that they fit into a single
unified picture. We note for instance that two PEGs (1241+735 and
2320+203) of the 200~mJy sample are completely indistinguishable from
BL Lacs not only in  pc-scale morphology, but also in the
level of polarization detected in the pc-scale jet (Bondi et al.,
2004). However, in this paper we find that, despite the
similarity in radio properties, the non-BL Lacs have on average
stronger emission lines than the FR1 radio galaxies. Furthermore, the
strong correlation between observed core strength and emission line
strength, both for the UGC sample alone and the UGC sample plus
200~mJy, is {\it prima facie} evidence for a tight relationship between
the intrinsic core strength and the emission lines that are somehow
excited by this core activity. Is it possible to fit all we know about
these objects into a single framework?

The simplest interpretation is to postulate an orientation
dependence of the emission lines which makes them appear
systematically stronger as the angle to the line of sight
decreases. This could go some way to account for why stronger emission
lines go with stronger cores but fails on three counts:

\begin{enumerate}

\item It does not explain why the BL Lacs, which are believed to be
 viewed at the smallest angles, do not have strong lines. 

\item It does not account for the very tight correlation between radio
core strength and line emission in the UGC sample which has been
selected on extended optical and radio properties, and  which should not
introduce any orientation bias.

\item The amount of extinction ($\sim 5^{m}$) required to restore the
correlation in Figure \ref{fig.lvlbi_lha} seems much too large when
compared with the recent measurements by Chiaberge (2002) for FR~I
radio galaxies.

\end{enumerate}

Any complete unified model must have an intrinsic correlation between
core strength and line emission built in from the start. This may be
possible. In a subsequent paper we will explore a model based on the
following ideas:

\begin{itemize}

\item Jets have a highly relativistic spine and a slower sheath (Laing
\& Bridle, 2004).

\item The relativistic spine of the jet is where the characteristic BL
Lac emission comes from and is usually only dominant when viewed at
small angles to the line of sight.

\item It is the energy dumped into the slow sheath that drives the
production of emission lines and produces the majority of the observed
core emission at large viewing angles.

\end{itemize}

\section{Summary and conclusions}

We summarize  our results as follows:

\begin{itemize}

\item We have compared the H$_{\alpha}$ emission properties of the 200~mJy
sample of low-luminosity core-dominated sources with those of a
matched sample of FR~I radio galaxies and find that the former have
significantly stronger emission lines. Either unified models based on
beaming have to be modified to include some orientation-dependent
extinction, or the dispersion in intrinsic core strengths is so large
that beaming is not the dominant factor. 

\item From combining observations of UGC galaxies with those of our
200~mJy sample we see that the observed core and emission line
strengths in radio galaxies are strongly correlated, with little
evidence for the kind of behaviour one might expect if the radio
emission from the cores of most of the core-dominated objects were 
beamed. The evidence for beaming amongst the many BL Lacs may be
strong but it is far from strong amongst the many objects with virtually
identical radio properties.

\end{itemize}

We conclude that FR~I unification is much more complex than usually
portrayed and models combining beaming with an intrinsic relationship
between core and emission line strengths need to be explored.

\section{Acknowledgments}
The National Radio Astronomy Observatory is a facility of the National
Science Foundation operated under cooperative agreement by Associated
Universities, Incorporated. We thank Marco Bondi and Antonis Polatidis
for allowing us to use some of their VLBI results prior to
publication. M.J.M. March\~a and S. Ant\'on acknowledge the financial
support of the Funda\c{c}\~ao para a Ci\^encia e Tecnologia through
the grants SFRH/BPD/3610/2000 and SFRH/BPD/5690/2001,
respectively. This research has made use of the NASA/IPAC
Extragalactic Database (NED) which is operated by the Jet Propulsion
Laboratory, California Institute of Technology, under contract with
the National Aeronautics and Space Administration.


\begin{table*}
\begin{center}
\caption{The 200mJy sample and the comparison sample. Cols: (1) Name;
  (2 and 11) Extended flux density at 20cm; (3 and 10) redhsift; (4)
  spectroscopic type: Sy1,2 stands for Sy-type 1,2; PEG stands for
  Passive Elliptical Galaxies; hyb stands for hybrid; BL stands for BL
  Lac; (5) log of line luminosity; (6) log of extendend luminosity;
  (7 and 12) equivalent width of $H_{\alpha}+[NII]$ in \AA; (8) SED type
  taken from Ant\'on et al., 2004: SPL for Steep Power-law, bPL for
  broken power-law and PL+IR concave spectrum with one or more bumps;
  (9) Name of twin (B1950). Flux densities
are in mJy, radio luminosities in W~Hz$^{-1}$ and line luminosities in
W.  }

{\tiny
\begin{tabular}{llllllllllll}
 \hline \hline
& & & & & & & &  & & &   \\
\multicolumn{1}{c}{Galaxy (B1950)} &  
\multicolumn{1}{c}{S$_{ext}$} & \multicolumn{1}{c}{z} & Type &  Log(L$_{\rm H_{\alpha}}$) & Log(L$_{ext}$) & EW  &  SEDs & twin & z & S$_{ext}$ &Log(L$_{\rm H_{\alpha}}$)  \\ 
& & & & & & & & & & & \\
\hline \hline
& & & & & & & & & & & \\
0046+316 &   10 & 0.015 & Sy2 &  33.8 & 21.6 & 150  &   PL+IR   &  0147+360  & 0.017 &  23 & 32.7 \\ 
0055+300 &  576 & 0.015 & PEG &  33.0 & 23.5 &  7  &   PL+IR   &  0123-016A & 0.018 & 910 & 32.8 \\ 
0125+487 &    6 & 0.067 & hyb &  34.0 & 22.8 & 47   &   bPL  & 0037+209  & 0.058 &  12 & $<$33.6\\ 
0149+710 &  239 & 0.022 & PEG &  33.1 & 23.4 & 18 &   PL+IR  &  1257+282  & 0.024 & 450 & $<$32.8\\
0210+515 &   48 & 0.049 & BL  & $\leq$33.3 & 23.4 & $\leq$3 &   bPL  &  0112-000  & 0.045 &  49 & 33.3\\ 
0321+340 &   98 & 0.061 & Sy1 &  35.0 & 23.9 & 200  &   bPL  & 0055-220 & 0.059 & 97 & $<$33.5\\ 
0651+410 &    9 & 0.021 & PEG &  33.6 & 21.9 & 26   &  PL+IR    & 1559+161  & 0.036 &  17 & 32.8\\ 
0806+350 &   31 & 0.082 & BL  & $\leq$33.5& 23.7    & $\leq$5      &  bPL  & 0135+185 & 0.072 & 26 & $<$33.8\\
0836+290 &  105 & 0.079 & PEG &  34.0 & 24.1 & 11    &   ?    & 0146+138 & 0.070 & 132 & $<$33.6\\
0848+686 &  163 & 0.0407& PEG &  33.8 & 23.7 & 10  &   SPL    &  1602+178A & 0.041 & 119 & 33.8 \\ 
1101+384 &   55 & 0.031 & BL  & $\leq$34.6 & 23.0 & $\leq$5  &   bPL & 1132+493  & 0.033 &  31 & $<$33.0 \\
1133+704 &  186 & 0.046 & BL  & $\leq$33.7 & 23.9 & $\leq$5  &   bPL  & 2322+143A & 0.045 & 187 & $<$33.5 \\
1144+352 &   29 & 0.063 & PEG &  34.5 & 22.4 & 25   &   PL+IR   & 1356+282  & 0.064 &  26 & 33.7 \\ 
1241+735 &  102 & 0.075 & PEG &  33.5 & 24.1 & 5    &  bPL   & 2336+212  & 0.072 & 145 & 33.6 \\ 
1646+499 &   41 & 0.0487& hyb &  34.3 & 23.3 & 75   &  bPL  & 2348+058  & 0.056 &  50 & 33.6 \\
1652+398 &   16 & 0.03  & BL  &  34.0 & 22.6 & 5  &   bPL  & 1130+493  & 0.031 &  30 & 33.9 \\
1703+223 &  173 & 0.049 & PEG & $\leq$33.1 & 23.9 & $\leq$5  & SPL   &  1250-150A & 0.045 & 210 & 33.1\\ 
1807+698 &  459 & 0.0505& BL  &  33.8 & 24.4 & 5    &  bPL  &  2154-080B & 0.057 & 430 & $<$33.2 \\
2116+81  &  110 & 0.084 & Sy1 &  36.1 & 24.2 & 557  &  bPL  &  2305+174 & 0.079 & 226 & $<$33.7 \\
2320+203 &  444 & 0.038 & PEG &  33.5 & 24.1 & 7   &   bPL   & 1132+492  & 0.032 & 475 & 33.0 \\ 

& & & & & &  & & & & &  \\
\hline \hline
\end{tabular}}
\end{center}

\label{obs.tab}
\end{table*}

\end{document}